\documentclass[sigconf]{acmart}

\AtBeginDocument{%
  \providecommand\BibTeX{{%
    \normalfont B\kern-0.5em{\scshape i\kern-0.25em b}\kern-0.8em\TeX}}}

\setcopyright{acmlicensed}
\copyrightyear{2023}
\acmYear{2023}
\acmDOI{10.1145/3580305.3599798}

\acmConference[KDD '23]{Proceedings of the 29th ACM SIGKDD Conference on Knowledge Discovery and Data Mining}{August 6--10,
  2023}{Long Beach, CA, USA}
 \acmBooktitle{Proceedings of the 29th ACM SIGKDD Conference on Knowledge Discovery and Data Mining (KDD '23), August 6--10, 2023, Long Beach, CA, USA}
\acmPrice{15.00}
\acmISBN{979-8-4007-0103-0/23/08}

\usepackage{algorithm}  
\usepackage{algpseudocode}  
\usepackage{multirow}
\usepackage{balance}

\begin{document}

\title{CT4Rec: Simple yet Effective Consistency Training for Sequential Recommendation}


\author{Chong Liu}
\authornote{Both authors contributed equally to this paper.}
\affiliation{%
  \institution{Tencent Inc.}
  \country{China}
}
\email{nickcliu@tencent.com}

\author{Xiaoyang Liu}
\authornotemark[1]
\affiliation{%
 \institution{OPPO Inc.}
 \country{China}
}
\email{liuxiaoyang@oppo.com}

\author{Rongqin Zheng}
\affiliation{%
  \institution{Tencent Inc.}
  \country{China}
  }
\email{leonezheng@tencent.com}

\author{Lixin Zhang}
\affiliation{%
  \institution{Tencent Inc.}
    \country{China}
  }
\email{lixinzhang@tencent.com}


\author{Xiaobo Liang}
\affiliation{%
  \institution{Soochow University}
    \country{China}
}
\email{xbliang3@stu.suda.cn}

\author{Juntao Li}
\authornote{Juntao Li is the corresponding author.}
\affiliation{%
 \institution{Soochow University}
  \country{China}
}
\email{ljt@suda.edu.cn}

\author{Lijun Wu}
\affiliation{%
 \institution{Microsoft Research Asia}
 \country{China}
}
\email{lijunwu@microsoft.com}

\author{Min Zhang}
\affiliation{%
 \institution{Soochow University}
  \country{China}
  }
\email{minzhang@suda.edu.cn}

\author{Leyu Lin}
\affiliation{%
  \institution{Tencent Inc.}
  \country{China}
  }
\email{goshawklin@tencent.com}

\renewcommand{\shortauthors}{Chong Liu et al.} 

\begin{abstract}
Sequential recommendation methods are increasingly important in cutting-edge recommender systems. 
Through leveraging historical records, the systems can capture user interests and perform recommendations accordingly.
State-of-the-art sequential recommendation models proposed very recently combine contrastive learning techniques for obtaining high-quality user representations. Though effective and performing well, the models based on contrastive learning require careful selection of data augmentation methods and pretext tasks, efficient negative sampling strategies, and massive hyper-parameters validation. In this paper, we propose an ultra-simple alternative for obtaining better user representations and improving sequential recommendation performance.
Specifically, we present a simple yet effective \textbf{C}onsistency \textbf{T}raining method for sequential \textbf{Rec}ommendation (CT4Rec) in which only two extra training objectives are utilized without any structural modifications and data augmentation. Experiments on three benchmark datasets and one large newly crawled industrial corpus demonstrate that our proposed method outperforms SOTA models by a large margin and with much less training time than these based on contrastive learning. Online evaluation on real-world content recommendation system also achieves 2.717\% improvement on the click-through rate and 3.679\% increase on the average click number per capita. 
Further exploration reveals that such a simple method has great potential for CTR prediction.
Our code is available at \url{https://github.com/ct4rec/CT4Rec.git}.
\end{abstract}






\maketitle

\section{Introduction}
Recommendation systems have been extensively applied in online platforms nowadays, e.g.,  Amazon \cite{amazon}, Google \cite{youtube_dnn,youtube1,NIS} and Facebook~\cite{ebr}. Due to the dynamic interactions between users and items, it is essential to capture evolving user interests from users’ historical records. To accurately represent user interests and make an appropriate recommendation, many efforts have been paid to study sequential recommendation methods \cite{Caser,sasrec,GRU4Rec,SR-GNN}.

Generally, the sequential recommendation task aims to characterize user representation from users' historical behaviors and predict the expected item accordingly. 
In viewing the great success of deep learning for sequential dependency modeling and representation learning, many methods based on deep neural networks \citep*{r1} have been introduced and proposed to solve this task,
covering RNN-Based frameworks~\cite{GRU4Rec,GRU4Rec+,STAMP}, different CNNs blocks and structures~\cite{Caser,3D_CNN}, Graph Neural Networks (GNNs) \cite{SR-GNN,FGNN,GC-SAN}, and also model variants \cite{sasrec,BERT4Rec,TiSASRec} relied on the powerful multi-head self-attention \cite{atten}. 
Though performing well, these methods might suffer from the data sparsity problem \cite{Autoint,s3,sasrec} for sequential recommendation, especially for models built on the multi-head self-attention mechanism, where only one single item prediction loss is used to optimize the full model parameters for capturing all possible correlations in input interaction sequences.

To address the above challenge, various self-supervised learning strategies are introduced~\cite{yao2020self,CLRec}.
Among these, the recently introduced contrastive learning (CL) objective~\cite{CL4SRec} achieves very promising results.
Through combing with effective data augmentation strategies and cooperating with the vanilla sequential prediction objective, the CL-based method can learn better sequence-level user representations and enhance the performance of sequential recommendations.
However, the effectiveness of CL-based approaches is subject to the correlated data augmentation methods, pretext tasks, efficient negative sampling, and hyper-parameters selection (e.g., the temperature in NCE and InfoNCE losses) \cite{jaiswal2021survey}. 
To mitigate the above drawbacks, many efforts have been made to simplify contrastive learning.
\cite{gao2021simcse} propose a very simple yet effective contrastive learning scheme that utilizes dropout noise as data augmentation to construct high-quality positive samples for sequence-level representation learning. 
Although such a scheme is effective for unsupervised representation learning, adapting it into the sequential recommendation task will still encounter the dilemma in which a higher proportion of CL objective in model training will lead to better representations that can easily distinguish positive and negative samples but might result in worse item prediction performance and vice versa.
In other words, there is a noticeable gap between the discrimination of positive and negative samples and the task objective for the sequential recommendation.
Thus, the key to obtaining better user representations mainly for the item prediction task is designing more training strategies that can address the inherent issues of sequential recommendation models.


Inspired by the recent observation on the multi-head attention model that a very simple regularization strategy imposed on the output space of supervised tasks yields striking performance improvement~\cite{liang2021r} (achieving SOTA on many challenging tasks), we propose to thoroughly explore the effect of consistency training for the sequential recommendation task.
We first introduce the simple bidirectional KL divergence regularization into the output space to constrain the inconsistency between two forward passes with different dropouts.
Unlike previous machine translation, summarization, and natural language understanding tasks, we argue that the introduced consistency regularization merely in the output space is not enough for the data sparsity setting of sequential recommendation.
We then design a novel and simple regularization objective in the representation space.
Unlike previous studies that utilize cosine \cite{gao2021simcse} and L2 \cite{ma2017dropout,zolna2018fraternal} distance to regularize the representation space, we propose to regularize the distributed probability of each user representation over others.
Thus, we can extend and leverage the effective bidirectional KL loss to regularize the representation inconsistency.
Experiments on three public benchmarks and one newly collected large-scale corpus indicate that our proposed simple consistency training for sequential recommendation (CT4Rec) outperforms state-of-the-art methods based on contrastive learning by a large margin.
Extensive experiments further prove that our proposed consistency training can be easily extended to the data side.
The Online A/B test also confirms the effectiveness of our method.
Besides, we extend our consistency training method to the CTR prediction task, and experiments conducted on two newly constructed industrial datasets further prove the effectiveness of our method.
In a nutshell, we mainly have the following contributions:
\begin{itemize}
\item We propose a simple (with only two bidirectional KL losses) yet very effective consistency training method for sequential recommendation systems.
To the best of our knowledge, this is the first work to thoroughly study the effect of consistency training from different perspectives and with a unified training objective for the sequential recommendation task.
\item Our proposed consistency training method can be easily extended to other inconsistency scenarios and tasks, e.g., data augmentation and CTR prediction. 
\item Extensive experiments on four offline datasets show the effectiveness of our proposed CT4Rec over SOTA models based on contrastive learning with much better performance and faster convergence time. The Online A/B test also shows significant improvement over the strong ensemble model.
\end{itemize}
\section{Related Work}
Early sequential recommendation (SR) methods are usually based on Markov Chain (MC), including adopting first-order MC \cite{FPMC} and high-order MCs \cite{TransRec,Fossil}. 
As for current deep learning approaches, they can be generally divided into four categories, i.e., RNN-based \cite{GRU4Rec,NARM,HRNN,GRU4Rec+,STAMP,jing2017neural,liu2016context,LSTM}, CNN-based \cite{Caser,3D_CNN,RCNN}, attention-based \cite{sasrec,BERT4Rec,TiSASRec,SHAN,MARank,ATRank} and GNN-based \cite{SR-GNN,FGNN,GC-SAN,GCN,APP-GE,wang2020make} methods. 
Concretely, GRU4Rec \cite{GRU4Rec} applies RNN to SR and many variants have been proposed based on GRU4Rec by adding data augmentation GRU4Rec+ \cite{GRU4Rec+}, hierarchical RNN \cite{HRNN} and attention module \cite{Air,NARM,SDM}.
However, RNN-based methods usually exhibit worse performance than CNN-based and attention-based methods for the data sparsity setting. 
Caser \cite{Caser} proposes a convolutional sequence embedding recommendation model, and \cite{3D_CNN} use 3-dimensional CNNs to achieve character-level encoding of input data.
Meanwhile, attention mechanisms \cite{atten} is used to model user behavior sequences and have achieved outstanding performance \cite{sasrec,BERT4Rec,TiSASRec,E-BART4Rec,fan2021lighter,liu2021augmenting}. Besides, many researches \cite{SR-GNN,FGNN,GC-SAN} combine attention mechanisms and GNNs to solve the SR task. 
Memory networks \cite{MANN,KV-MN}, data augmentation \cite{GLaS,SSL} and session-based \cite{SWIWO)} models are also utilized to improve the performance of SR.

Recently, the combination of contrastive learning and attention mechanisms \cite{CL4SRec,CLRec} is widely used in SR and achieves great success.
CauseRec \cite{CauseRec} performs contrastive learning by contrasting the counterfactual with the observational.
StackRec \cite{StackRec} utilizes stacking and fine-tuning to develop a deep but easy-to-train SR model. 
ICAI-SR \cite{ICAI-SR} focuses on the complex relations between items and categorical attributes in SR. 
Unlike previous researches, we only introduce two consistency training objectives in the representation and output spaces without any structure modifications, extra data, and heuristic patterns from tasks.
Though extremely simple, our method outperforms recent SOTA models by a large margin and is more efficient for model training.

\begin{figure*}[th]
    \centering
    \includegraphics[width=1\textwidth]{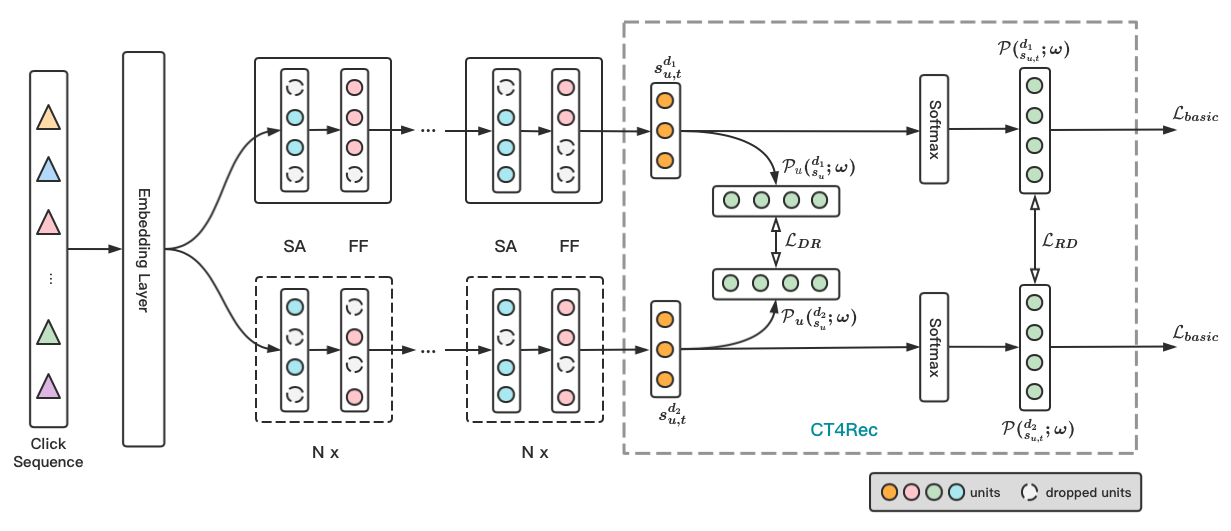}
    \caption{Model structure of CT4Rec. It takes user click sequences as input and outputs user representations for item retrieval in the matching stage of recommendation. The input sequences are transformed into vector representations via the embedding layer and then encoded by N transformers with different hidden dropout masks. In addition, Distributed Regularization Loss and Regularized Dropout Loss are introduced to restrain these representations generated by different dropout masks.}  \label{CT4Rec}
    \end{figure*} 
    
There are only a few related researches in the fields of natural language process~\cite{gao2021simcse} and machine learning~\cite{ma2017dropout,zolna2018fraternal,liang2021r}.
Specifically, \cite{gao2021simcse} introduce dropout as the alternative of data augmentation into contrasting learning for sequence representation learning while we focus on the inconsistency introduced by dropout in the data sparsity SR task.
Our proposed method is also extremely simple compared with the paradigm of combining contrastive learning with the traditional item prediction objective.
\cite{ma2017dropout,zolna2018fraternal} mainly focus on the gap between training and testing and utilize L2 for regularizing the representation space, which is less effective in the data sparsity setting, represented by the marginal to none performance improvements in Section~\ref{sec:analysis}.
Different from introducing a regularization objective in the output space to constrain the randomness of sub-models brought by dropout \cite{liang2021r}, we focus on the consistency training of the data sparsity SR task from both the representation and output space.
We also propose a simple yet effective regularization strategy in the representation space to compensate and align the output space consistency loss.

\section{The CT4Rec Model}
\label{sec:method}
The overall structure of our model is illustrated in Figure \ref{CT4Rec}.
Before elaborating our proposed CT4Rec, we first present some necessary notations to describe the sequential item prediction task. Let \(\mathcal{U}=(u_1,u_2,...,u_{|\mathcal{U}|})\) denote a set of users, and \(\mathcal{V}=(v_1,v_2,...,v_{|\mathcal{V}|})\) denote a set of items.
The sequence for user \(u\in\mathcal{U}\) is denoted as \(s_u=(v_1^{(u)},v_2^{(u)},...,v_t^{(u)},...,v_{|s_u|}^{(u)})\), where \(v_t^{(u)}\in\mathcal{V}\) is the item that user $u$ interacts at time step $t$ and \(|s_u|\) is the length of sequence \(s_u\). Given the historical sequence \(s_u\), the task of sequential recommendation is to predict the probability of all alternative items to be interacted by user $u$ at time step \(|s_u|+1\), which is formulated as ${P}(v_{|s_u|+1}^{(u)}=v|s_u)$.

\subsection{Backbone Model}
Since our proposed consistency training method does not involve structural modification and extra data utilization, we apply it to the widely used SASRec model \cite{sasrec}.
Following the original setting in SASRec, the transformer encoder contains three parts, i.e., an embedding layer, the stacked multi-head self-attention blocks, and a prediction layer. Then, we can obtain user representation \(\boldsymbol{s_u}=f(s_u)\), where \(f(\cdot)\) indicates the transformer encoder.

To learn the relation between users and items in sequential recommendation, a similarity function, e.g., inner product, is applied to measure distances between user representation and item representation. Thus, for user representation \(\boldsymbol{s_{u,t}}\) of user $u$ at time step $t$, we can get a similarity distribution \(\mathcal{P}(\boldsymbol{s_{u,t}})=\mathcal{P}(v_{t+1}^{(u)}|\boldsymbol{s_{u,t}})\) to predict the item that user $u$ will interact at time step $t+1$. Then, the basic loss function with positive item \(\boldsymbol{ v_{t+1}^+}\) and randomly sampled negative items \(\boldsymbol{ v_{t+1}^-}\in{\mathcal{V}}\) is denoted as:
\begin{equation}
    \begin{split}
    \mathcal{P}(\boldsymbol{s_{u,t}};\omega) = 
    \frac{exp(\boldsymbol{s_{u,t} v_{t+1}^+})}{exp(\boldsymbol{s_{u,t} v_{t+1}^+}) +  \sum\limits_{\boldsymbol{v_{t+1}^-}\in{\mathcal{V}}}exp(\boldsymbol{s_{u,t}v_{t+1}^-}) }
    \label{func:basic_pro}
    \end{split}
\end{equation}
\begin{equation}
    \begin{split}
    \mathcal{L}_{basic}(\boldsymbol{s_{u,t}};\omega)= -log\mathcal{P}(\boldsymbol{s_{u,t}};\omega) \label{func:basic_loss}
    \end{split}
\end{equation}
, where $\omega$ refers to all trainable parameters of the model.
\begin{figure*}[t]
    \centering
    \includegraphics[scale=0.35]{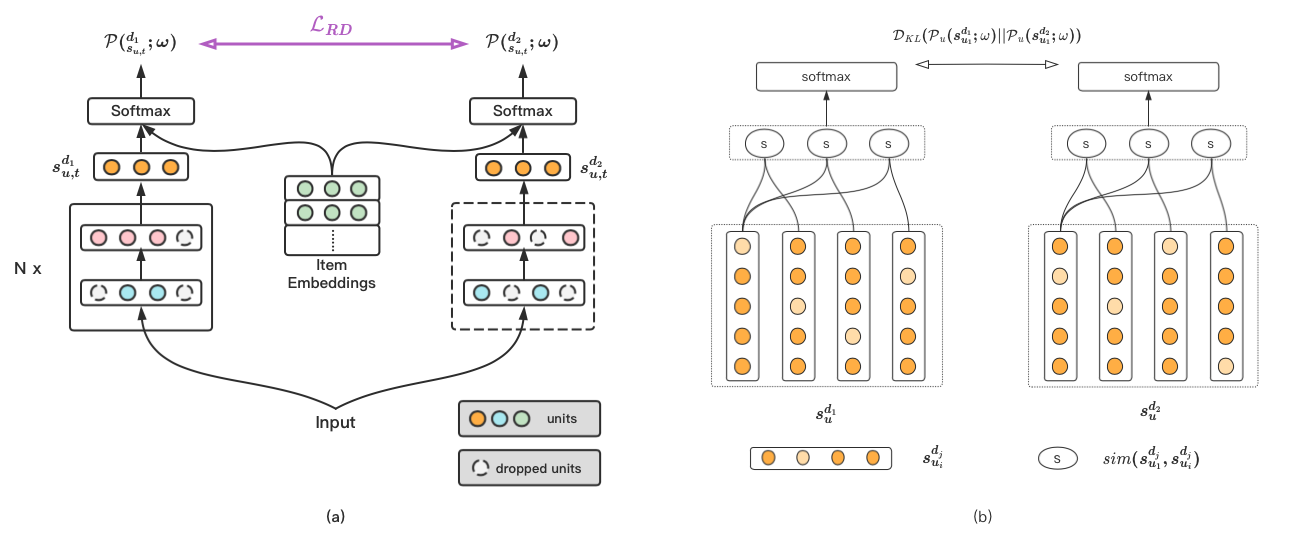}
    \caption{Illustration of (a) RD loss and (b) DR loss.
    }  \label{Fig:RD_DR}
    \end{figure*} 
\subsection{Consistency Training}
Since the over-fitting problem extensively exists in deep neural network models, regularization methods, including dropout, are widely used to alleviate this problem. Commonly, dropout can reduce over-fitting and co-adapting by randomly removing a certain rate of units in the whole deep neural network. Also, dropout can be treated as a method to generate and combine exponentially sub-models, which always effectively enhances model performance. Considering the above advantages and the randomness of dropout, we propose our CT4Rec based on dropout to regularize both the output space and the representation space of models.
Inspired by recent studies on dropout \cite{liang2021r}, we enhance the user representation from the perspective of reducing the model inconsistency and gap between training and testing.

Concretely, we forward twice with different dropouts and learn the consistency between these two representations for each user, i.e., each user interaction sequence \(\mathbf{s_u}\) passing the forward network twice and obtain two representations \(\boldsymbol{s_{u,t}^{d_1}}\) and \(\boldsymbol{s_{u,t}^{d_2}}\).
Since dropout randomly removes units in a model, the two representations are actually generated from two sub-models of the same model.

\textbf{Regularized Dropout Loss (RD).} We first apply a regularized dropout loss to constrain the output space of sub-models from dropout. Considering two representations \(\boldsymbol{s_{u,t}^{d_1}}\) and \(\boldsymbol{s_{u,t}^{d_2}}\) for user $u$, as mentioned in Function \ref{func:basic_pro}, we can get two similarity distributions \(\mathcal{P}(\boldsymbol{s_{u,t}^{d_1}};\omega)\) and \(\mathcal{P}(\boldsymbol{s_{u,t}^{d_2}};\omega)\). 
Then, we introduce a bidirectional KL-divergence loss to regularize the above two distributions:
\begin{equation}
    \begin{split}
    \mathcal{L}_{RD}(\boldsymbol{s_{u,t}};\omega)=\frac{1}{2}(\mathcal{D}_{KL}(\mathcal{P}(\boldsymbol{s_{u,t}^{d_1}};\omega)||\mathcal{P}(\boldsymbol{s_{u,t}^{d_2}};\omega))\\
    +\mathcal{D}_{KL}(\mathcal{P}(\boldsymbol{s_{u,t}^{d_2}};\omega)||\mathcal{P}(\boldsymbol{s_{u,t}^{d_1}};\omega))) \label{func:RD_loss}
    \end{split}
\end{equation}


As shown in Figure \ref{Fig:RD_DR}(a), the dropped units of the left model to generate user presentation \(\boldsymbol{s_{u,t}^{d_1}}\) are different from that of the right model to generate \(\boldsymbol{s_{u,t}^{d_2}}\). Therefore, the similarity distributions \(\mathcal{P}(\boldsymbol{s_{u,t}^{d_1}};\omega)\) and \(\mathcal{P}(\boldsymbol{s_{u,t}^{d_2}};\omega)\) are also varied for the same input sequence \(\mathbf{s_u}\).

\textbf{Distributed Regularization Loss (DR).} 
To better regularize the representation space, we propose a distributed regularization method in which each user is represented by its correlations with other users rather than directly utilizing user representations for consistency regularization.
In this paper, we compare users in each mini-batch, e.g., $n$ users \((u_1,u_2,...,u_n)\) and two representations for each user generated by dropout denoted as \((\boldsymbol{s_{u_1}^{d_1}}, \boldsymbol{s_{u_2}^{d_1}},..., \boldsymbol{s_{u_n}^{d_1}})\) and \((\boldsymbol{s_{u_1}^{d_2}}, \boldsymbol{s_{u_2}^{d_2}}, \dots, \boldsymbol{s_{u_n}^{d_2}})\). As shown in Figure \ref{Fig:RD_DR}(b), for user \(u_1\), we calculate the similarities between \(\boldsymbol{s_{u_1}^{d_j}}\) and all the other user representation \(\boldsymbol{s_{u_i}^{d_j}}\), defined as \(sim(\boldsymbol{s_{u_1}^{d_j}},\boldsymbol{s_{u_i}^{d_j}})\), and obtain the similarity distribution
\begin{equation}
    \begin{split}
    \mathcal{P}_{u}(\boldsymbol{s_{u_1}^{d_j}};\omega)=softmax(sim(\boldsymbol{s_{u_1}^{d_j}},\boldsymbol{s_{u_2}^{d_j}}),...,sim(\boldsymbol{s_{u_1}^{d_j}},\boldsymbol{s_{u_n}^{d_j}}))
      \label{func:DR_pro}   
    \end{split}
\end{equation}

Then, a bidirectional KL-divergence loss is applied to regularize the two distributions \(\mathcal{P}_{u}(\boldsymbol{s_{u_1}^{d_1}};\omega)\) and \(\mathcal{P}_{u}(\boldsymbol{s_{u_1}^{d_2}};\omega)\), defined as:
\begin{equation}
    \begin{split}
    \mathcal{L}_{DR}(\boldsymbol{s_u};\omega)=\frac{1}{2}(\mathcal{D}_{KL}(\mathcal{P}_u(\boldsymbol{s_u^{d_1}};\omega)||\mathcal{P}_u(\boldsymbol{s_u^{d_2}};\omega)) \\
    +\mathcal{D}_{KL}(\mathcal{P}_u(\boldsymbol{s_u^{d_2}};\omega)||\mathcal{P}_u(\boldsymbol{s_u^{d_1}};\omega)))
      \label{func:DR_loss}
    \end{split}
\end{equation}

    

\textbf{Final Objective.} 
As shown in Figure \ref{CT4Rec}, we train the above two objectives together with the task-specific loss in the backbone model.
The task-specific loss and final training objective are as:
\begin{equation}
    \begin{split}
    \mathcal{L}_{basic}(\boldsymbol{s_{u,t}};\omega)= -\frac{1}{2}( log\mathcal{P}(\boldsymbol{s_{u,t}^{d_1}};\omega)+ log\mathcal{P}(\boldsymbol{s_{u,t}^{d_2}};\omega)) \label{func:bs2}
    \end{split}
\end{equation}
\begin{equation}
    \mathcal{L}_{final}= \mathcal{L}_{basic} + \alpha \mathcal{L}_{RD} +\beta \mathcal{L}_{DR} \label{func:final_loss}
\end{equation}
where \(\alpha\) and \(\beta\) are the coefficient weights to control \(\mathcal{L}_{RD}\) and \(\mathcal{L}_{DR}\). Thus, our CT4Rec can control the influence of dropout and constrain the model space. Compared with Equation~\ref{func:basic_loss}, our CT4Rec only adds two losses \(\mathcal{L}_{RD}\) and \(\mathcal{L}_{DR}\) with model structures unchanged, which can also be widely applied on various model structures.
We present the training algorithm in the Appendix~\ref{sec:app1}.

\section{Experiments}\label{sec:exp}
\subsection{Datasets}
We conduct extensive experiments on three public benchmark datasets that are widely used in recent literature \cite{CL4SRec} and a new large-scale dataset collected from a real-world recommendation scenario, i.e., PC-WeChat Top Stories. 
These datasets are very different in domains, platforms, and data scale, where their detailed statistics are presented in Table \ref{tb1}.
\begin{table}[th]
    \caption{Dataset statistics of three public benchmarks and an offline corpus from real-world application, where \textit{avg.} refers to the average actions per user.}
    \begin{tabular}{cccccc}
    \toprule
         Dataset & \#users & \#items & \#actions & avg. & density \\
         \hline
         Beauty  & 52,024 & 57,289 & 0.4M & 7.6 & 0.01\% \\
         Sports & 25,598 & 18,357 & 0.3M & 8.3  & 0.05\%\\
         Yelp &  30,431 & 20,033  & 0.3M & 10.4  & 0.05\%  \\
         WeChat & 749,452 & 211,004 & 9.5M & 12.7 & 0.006\% \\  
\bottomrule     
    \end{tabular}
    \label{tb1}
    \vspace{-0.3cm}
\end{table}
\begin{itemize}
\item \textbf{Amazon:} 
a series of datasets comprise product reviews, which are crawled from one of the largest E-Commerce platforms, i.e., \textit{Amazon.com}.
As introduced in SASREC \cite{sasrec,image_dataset}, these datasets are separated by top-level product categories.
We follow one of the most recent researches \cite{CL4SRec} to utilize the \textbf{\textit{Beauty}} and \textbf{\textit{Sports}} categories for comparison. 

\item \textbf{Yelp:} it is a widely acknowledged dataset for the business recommendation, which is collected from the Yelp platform\footnote{https://www.yelp.com/dataset}. Following \cite{s3}, we leverage the data after January 1st, 2019, and treat business categories as attributes.
   
\item \textbf{WeChat:} 
this dataset is constructed from WeChat platform for PC Top Stories recommendation (denoted as \textbf{\textit{WeChat}} to distinguish datasets from other platforms), which consists of interaction records from 7th to 13rd, June 2021.
Each interaction provides positive feedback (i.e., click) of an item from a user. 
Concretely, we collect 9.5 million interactions from 0.74 million users on 0.21 million items and regard data from the first few days as the train set and the rest for testing.

\end{itemize}

\subsection{Baselines}
To verify the effectiveness of our proposed method, we introduce four representative baselines and three very recent methods.

\begin{itemize}
 \item \textbf{GRU4Rec \cite{GRU4Rec}.} It utilizes GRU modules to model user action sequences for the session-based recommendation. We consider each user's click sequence as a session.
 
\item \textbf{SASRec \cite{sasrec}.} It applies the multi-head self-attention mechanism to solve the sequential recommendation task, which is commonly treated as one of the state-of-the-art baselines.
In this paper, we utilize SASRec as the backbone of our CT4Rec.

\item \textbf{TiSASRec \cite{TiSASRec}.} Based on SASRec, it further introduces time interval aware self-attention mechanism to encode the user's interaction sequence, where the positions and time interval between any two items are considered.


\item \textbf{BERT4Rec \cite{BERT4Rec}.} 
It uses the deep bidirectional self-attention mechanism to model user interaction history in the sequential recommendation and trains the model like BERT \cite{devlin2019bert}.
 
\item \textbf{CL4SRec \cite{CL4SRec}.} It generates different views of the same user interaction sequence by using data augmentation methods and adds contrastive learning objective to the original objective of SASRec for sequential recommendation tasks.

\item \textbf{CLRec.} \cite{CLRec} design a queue-based contrastive learning method named CLRec to de-bias deep candidate generation in the recommendation system and further propose Multi-CLRec for multi-intention aware bias reduction. In this work, we only compare the CLRec for fairness.

\item \textbf{StackRec \cite{StackRec}.} It first uses a stacking operation on the pre-trained layers/blocks to transfer knowledge from a shallow model to a deep model and then utilizes iterative stacking to obtain a deeper but easier-to-train recommendation model.

\end{itemize}

\subsection{Settings}
All models are implemented based on TensorFlow. 
For baselines with official codes, we utilize the implementations provided by authors.
As for models without open-accessible codes from the original paper, we prefer the well-tested version from the open-source community.
Specifically, we use code from \url{https://github.com/Songweiping/GRU4Rec\_TensorFlow} as the implementation of GRU4Rec \cite{GRU4Rec}.
We implement CL4SRec and CLRec based on the model descriptions and experimental settings of the correlated papers since there is no official code or popular implementation from the open-source community.
For fair comparisons, the embedding dimension size is set to 50, and all models are optimized by Adam.

Recall that our proposed CT4Rec method does not modify the model architecture and increases the model scale of the backbone model.
Instead, it only involves two effective consistency regularization strategies.
Thus, we follow the backbone method, i.e., SASRec \cite{sasrec}, to implement our CT4Rec.
We use two self-attention layers and set the head number to 2. The maximum sequence length is 50 for all datasets. 
We optimize the parameters with the learning rate of 0.001 and the batch size as 128.
The dropout rate of turning off neurons is set to 0.5. 
To verify the effect of each component and their combination, we fix the structure and other hyper-parameters of the model and only adjust the values of $\alpha$ and $\beta$, 
where $\alpha$ and $\beta$ are selected from $\{0.1,0.3,0.5,1.0,2.0,3.0\}$. 
Other details for reproducing our experiments can be found in our anonymous code.

\subsection{Offline Evaluation}
\subsubsection{Evaluation Protocols.}
Following many previous studies \cite{r1,sasrec,s3,CL4SRec}, we employ the leave-one-out strategy to evaluate model performance. Specifically, for each user, we take the last interacted item for test. 
Similar to \cite{sasrec,s3}, we randomly sample 500 items from the whole dataset for each positive item, and rank them by similarity scores. 
The model performances are evaluated by top-k Normalizedrand Discounted Cumulative Gain (NDCG\({@k}\)) and top-k Hit Ratio (HR\({@k}\)), which are both commonly used in top-k recommendation systems. 
Specifically, we report HR\({@k}\) and NDCG\({@k}\) with $k=\left\{5,10,20\right\}$ for all datasets.
\subsubsection{Experimental Results.}
\begin{table*}[th]
\caption{Model performance of baselines and our proposed CT4Rec on four offline datasets, where `*' refers to modifying the original binary cross-entropy loss in SASRec with the training objective in recent baselines, e.g., CLRec, CL4SRec, StackRec. Our CT4Rec is implemented on SASRec*. 
\textit{Improv.} and \textit{Improv.*} refer to the relative improvement of CT4Rec over SASRec and SASRec*, respectively. The performance improvement over baselines is statistically significant with $p\textless 0.01$, in which we present the experimental results of extra two runs with different random seeds in the Appendix~\ref{sec:app2}.}
\renewcommand\arraystretch{1.06}
\resizebox{\textwidth}{!}{
\begin{tabular}{lllllllllllll}
\toprule  
Datasets   & Metric & SASRec & SASRec* & GRU4Rec & BERT4Rec & TiSASRec & StackRec & CLRec  & CL4SRec & CT4Rec  & Improv. & Improv.* \\
\hline   
\multirow{6}*{Beauty}& HR@5    & 0.2109 & 0.2194  & 0.1179  & 0.0860   & 0.2024   & 0.1725   & 0.1498 & 0.2275  & \textbf{0.2556} & 21.19\% & 16.50\%  \\
       & HR@10   & 0.2759 & 0.2748  & 0.1582  & 0.1357   & 0.2746   & 0.2225   & 0.1816 & 0.2896  & \textbf{0.3200} & 15.98\% & 16.45\%  \\
        & HR@20   & 0.3546 & 0.3392  & 0.2146  & 0.2036   & 0.3508   & 0.2818   & 0.2218 & 0.3624  & \textbf{0.3891} & 9.73\%  & 14.71\%  \\
& NDCG@5  & 0.1523 & 0.1661  & 0.0869  & 0.0572   & 0.1413   & 0.1286   & 0.1177 & 0.1701  & \textbf{0.1924} & 26.33\% & 15.83\%  \\
       & NDCG@10 & 0.1733 & 0.1840  & 0.1000  & 0.0732   & 0.1646   & 0.1447   & 0.1279 & 0.1901  & \textbf{0.2132} & 23.02\% & 15.87\%  \\
       & NDCG@20 & 0.1932 & 0.2003  & 0.1141  & 0.0903   & 0.1839   & 0.1596   & 0.1380 & 0.2085  & \textbf{0.2307} & 19.41\% & 15.18\%  \\
       \hline   
\multirow{6}*{Sports}        & HR@5    & 0.1912 & 0.1966  & 0.0961  & 0.0766   & 0.1703   & 0.1318   & 0.1498 & 0.2084  & \textbf{0.2196} & 14.85\% & 11.70\%  \\
       & HR@10   & 0.2747 & 0.2683  & 0.1499  & 0.1267   & 0.2456   & 0.1962   & 0.2178 & 0.2834  & \textbf{0.3010} & 9.57\%  & 12.19\%  \\
       & HR@20   & 0.3751 & 0.3547  & 0.2295  & 0.2055   & 0.3352   & 0.2780   & 0.3072 & 0.3721  & \textbf{0.3950} & 5.31\%  & 11.36\%  \\
       & NDCG@5  & 0.1289 & 0.1398  & 0.0631  & 0.0494   & 0.1159   & 0.0902   & 0.1044 & 0.1488  & \textbf{0.1556} & 20.71\% & 11.30\%  \\
       & NDCG@10 & 0.1558 & 0.1629  & 0.0804  & 0.0654   & 0.1402   & 0.1110   & 0.1263 & 0.1729  & \textbf{0.1817} & 16.62\% & 11.54\%  \\
       & NDCG@20 & 0.1811 & 0.1847  & 0.1003  & 0.0852   & 0.1628   & 0.1315   & 0.1488 & 0.1953  & \textbf{0.2055} & 13.47\% & 11.26\%  \\
       \hline   
\multirow{6}*{Yelp}& HR@5    & 0.2834 & 0.3216  & 0.1457  & 0.1567   & 0.2935   & 0.2230   & 0.2545 & 0.3173  & \textbf{0.3462} & 22.16\% & 7.65\%   \\
       & HR@10   & 0.4221 & 0.4469  & 0.2546  & 0.2623   & 0.4257   & 0.3397   & 0.3881 & 0.4451  & \textbf{0.4784} & 13.34\% & 7.05\%   \\
       & HR@20   & 0.5975 & 0.5989  & 0.4257  & 0.4312   & 0.5839   & 0.4927   & 0.5670 & 0.5993  & \textbf{0.6309} & 5.59\%  & 5.34\%   \\
       & NDCG@5  & 0.1889 & 0.2283  & 0.0890  & 0.0996   & 0.2009   & 0.1484   & 0.1702 & 0.2236  & \textbf{0.2443} & 29.33\% & 7.01\%   \\
       & NDCG@10 & 0.2335 & 0.2687  & 0.1239  & 0.1336   & 0.2435   & 0.1859   & 0.2132 & 0.2647  & \textbf{0.2869} & 22.87\% & 6.77\%   \\
       & NDCG@20 & 0.2778 & 0.3070  & 0.1668  & 0.1759   & 0.2834   & 0.2244   & 0.2582 & 0.3036  & \textbf{0.3253} & 17.10\% & 5.96\%   \\
       \hline  
\multirow{6}*{WeChat}  &HR@5    & 0.2756 & 0.3069   & 0.1836  & 0.1943   & 0.3193   & 0.2975   & 0.2849 & 0.3105 & \textbf{0.3406} & 25.58\% & 10.98\%   \\
 &HR@10   & 0.4103 & 0.4366   & 0.2231  & 0.2247   & 0.4406   & 0.4173   & 0.3985 & 0.4511 & \textbf{0.4861} & 18.47\% & 11.34\%   \\
 &HR@20   & 0.5291 & 0.5484   & 0.2884  & 0.2907   & 0.5539   & 0.5197   & 0.4852 & 0.5507 & \textbf{0.5979} & 13.00\% & 9.03\%  \\
& NDCG@5  & 0.1948 & 0.2131   & 0.1272  & 0.1266   & 0.2036   & 0.2082   & 0.1939 & 0.2195 & \textbf{0.2361} & 21.20\% & 10.79\%   \\
 &NDCG@10 & 0.2357 & 0.2743 & 0.1447 & 0.1394 & 0.2615 & 0.2576 & 0.2371 & 0.2827 & \textbf{0.3057} & 29.70\% & 11.40\% \\
 &NDCG@20 & 0.2869 & 0.3013   & 0.1693  & 0.1526   & 0.2957   & 0.2811   & 0.2639 & 0.3089 & \textbf{0.3314} & 15.51\% & 9.99\%    \\
\bottomrule
\end{tabular}}
\label{tb2}
\end{table*}

As shown in Table~\ref{tb2}, our proposed CT4Rec outperforms all other baseline methods, including multiple representative models and state-of-the-art sequential recommendation solutions, on three benchmark datasets and one large industrial corpus. 
Compared with other strong methods that utilize data augmentation or/and contrastive learning to obtain better user representations and mitigate the incompatibility between the single item prediction task and the vast amounts of parameters in data sparsity scenarios \cite{sasrec}, our CT4Rec is simpler and more effective without requiring any augmented data or delicate training strategies in which only two extra objectives are introduced.
The performance advance of CT4Rec over SOTA models based on contrastive learning is confirmed by the significant improvements of HR\({@k}\), and NDCG\({@k}\) scores over CLRec and CL4SRec.
We also observe that the training objective in recent baselines performs much better than the original binary cross-entropy in \textit{SASRec}, which is demonstrated by the results that \textit{SASRec*} increase offline scores by a large margin over \textit{SASRec}.
Notice that our CT4Rec is implemented by introducing two training objectives into \textit{SASRec*}.
We compare \textit{CT4Rec} with \textit{SASRec*} to calibrate the effect of our consistency training method.
The universal enhancement of \textit{CT4Rec} over \textit{SASRec*} on four datasets and all HR\({@k}\) and NDCG\({@k}\) (relative improvements ranging from 5.34\% to 16.50\%) verify the superiority of our proposed simple method. 


\subsection{Online Evaluation}
\subsubsection{Evaluation Protocols.}
We further perform an online A/B test that resembles \cite{xie-AFT} to evaluate our CT4Rec in a real-world system. 
Commonly, an online recommendation system is divided into four stages, including matching, pre-ranking, ranking, and re-ranking.
We have deployed CT4Rec for \textit{PC Wechat Top Stories} recommendation in the matching stage.
To calibrate the effect of CT4Rec for the online system, CT4Rec is implemented as an additional channel in the current matching module with the rest of the system unchanged, where the existing matching module refers to an ensemble model that combines multiple methods, e.g., rule-based, reinforcement-based, sequence-based, DSSM, self-distillation.
In the online A/B test, we utilize two metrics, i.e., click-through rate (CTR) and average click number per capita (ACN), to evaluate model performance. 
The online test lasts for 7 days and involves nearly 3 million users.
\begin{table}[]
    \caption{Performance improvement of online A/B test.}
    \centering
    \begin{tabular}{cc}
        \toprule
          CTR & ACN   \\
         \hline
         +2.717\%   & +3.679\%  \\
        \bottomrule     
    \end{tabular} 
    \label{tb4}
\end{table}
\subsubsection{Experimental Results.}
The performance improvement of CT4Rec on real-world recommendation service is reported in Table \ref{tb4} with the significance level $p<0.01$, which can be seen that only implementing CT4Rec as an additional channel in the matching module with the rest of the system unchanged can yield very impressive performance improvement over the original ensemble model.
Our method increases CTR and ACN metrics by 2.717\% and 3.679\%, respectively.
This indicates that our proposed simple method is not only effective for offline benchmarks and evaluation metrics but also generalizes well to the real-world online system.
Since CT4Rec restricts the uncertainty of sub-models without involving extra model structure, the computation cost of CT4Rec for online serving is identical to the original backbone model.
Thus, CT4Rec can significantly enhance online performance without adding online computation costs, which is a definite advantage for online serving, especially considering the machine costs.

\section{Analysis}
\label{sec:analysis}
\begin{figure*}[htbp]
    \centering
    \includegraphics[width=1\textwidth]{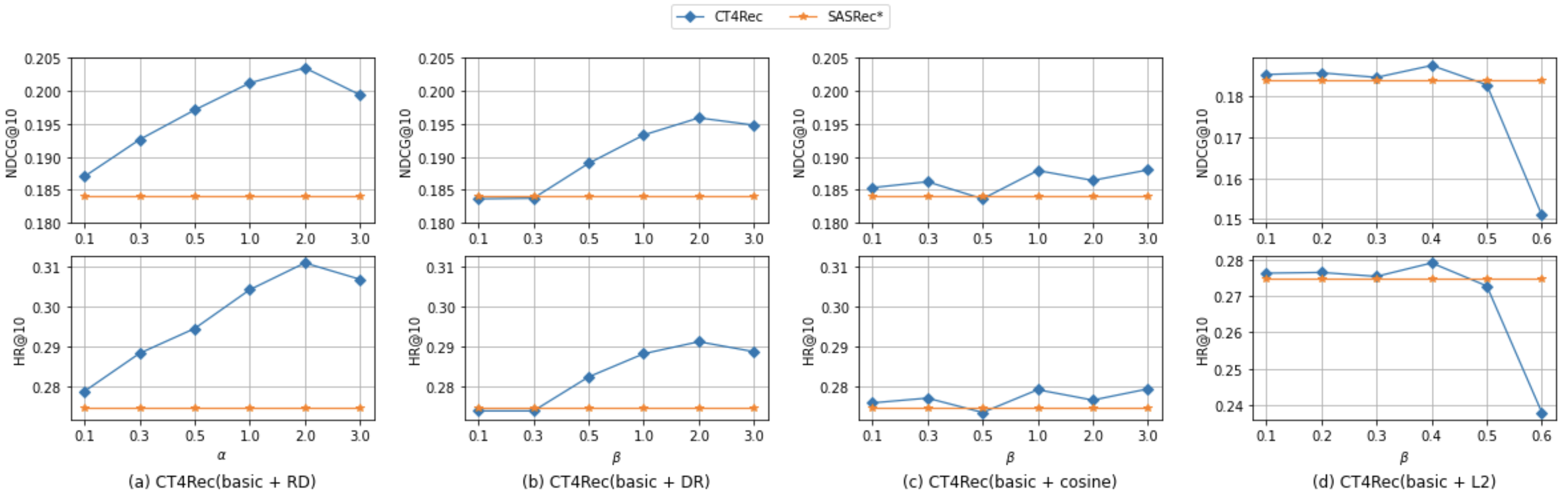}
    \caption{Evaluation results for ablation study and analysis of $\alpha$ and $\beta$. (a) CT4Rec with $\mathcal{L}_{basic}$ and $\mathcal{L}_{RD}$. (b) CT4Rec with $\mathcal{L}_{basic}$ and $\mathcal{L}_{DR}$. (c) and (d) replace $\mathcal{L}_{DR}$ with cosine and L2 loss, respectively, where the orange lines are the performance of SASRec*.}  \label{img:ablation}
    \end{figure*} 
We have demonstrated the impressive performance of our proposed simple CT4Rec on offline benchmarks and the online A/B test.
In this section, we launch extensive experiments to understand and analyze CT4Rec from different perspectives. 
For convenience, these studies are performed on \textit{Beauty}. 
More specifically, we mainly focus on: \textbf{1)} the effect of each introduced objective (\textit{Ablation Study}), \textbf{2)} the influence of several important hyper-parameters (\textit{Hyper-Parameter Analysis}), 
\textbf{3)} the extension of CT4Rec to data augmentation (\textit{Extension to Data Augmentation}),
\textbf{4)} the change of training process and cost resulted by CT4Rec (\textit{Training and Cost Analysis})
\subsection{Ablation Study}    
We perform an ablation study to explore the effect of two objectives in Section~\ref{sec:method}, i.e., regularized dropout (RD) and distributed regularization (DR), where the results are illustrated in Figure~\ref{img:ablation}.

\textbf{RD Objective.} As shown in the left sub-figure (a), the introduced RD objective achieves significant performance improvement over the backbone \textit{SASRec*} model.
Compared with DR objective and its variants, RD loss contributes most to the performance increase, which concludes that launching consistency regularization in the model output space is the most beneficial to \textit{SASRec*}, which is similar to the observations in other tasks~\cite{liang2021r}.
The possible reason might be that RD consistency regularization affects directly on the same space of the model output probability distribution.

\textbf{DR Objective.} We can see that the unsupervised DR loss can also yield substantial performance gains, which points out that unsupervised consistency regularization in user representation space for data sparsity sequential recommendation task is also essential.
We also adapt two typical unsupervised strategies in previous studies \cite{ma2017dropout,zolna2018fraternal,gao2021simcse} to regularize user representations, including cosine similarity and L2 distance.
As shown in the right two sub-figures in Figure~\ref{img:ablation}, these two methods have not brought meaningful improvement upon the backbone method, which further proves that our designed DR objective is more preferable for consistency regularization in the representation space.


    
\begin{figure}[th]
    \centering
    \includegraphics[width=0.45\textwidth]{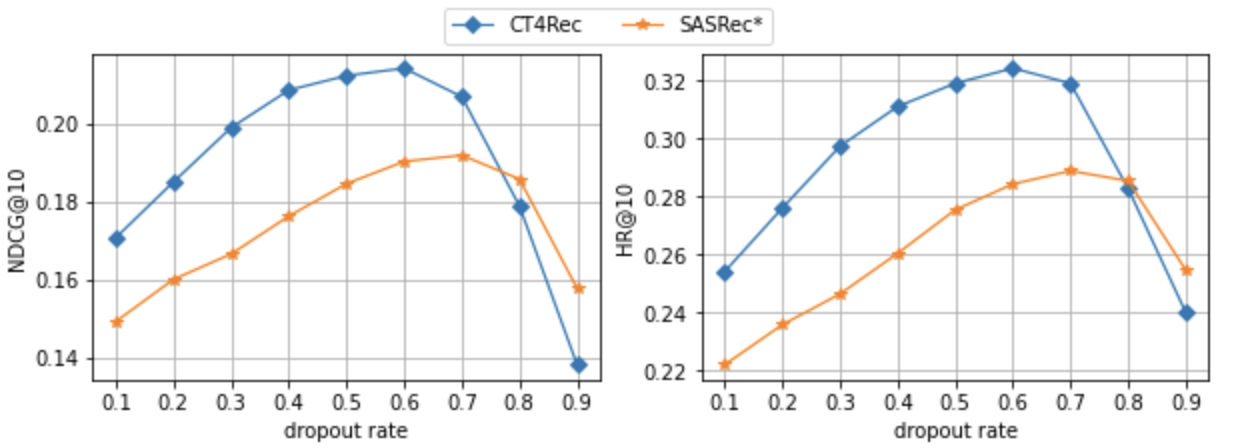}
    \caption{The impact of different dropout rates for CT4Rec and SASRec* on the Beauty dataset.}  \label{img:dropout}
    \vspace{-0.25cm}
    \end{figure} 
\subsection{Hyper-Parameter Analysis}
We mainly consider several critical hyper-parameters in this part, including $\alpha$ and $\beta$ in Equation~\ref{func:final_loss}, and the dropout rate.
\begin{table*}[t]
\caption{Performance comparison of different consistency regularization strategies in the data augmentation setting.}\label{tab:data_augmentation}
\resizebox{\textwidth}{!}{
\begin{tabular}{c|c|cc|ccc|ccc|r r}
\toprule
Aug.\cite{CL4SRec} & Metrics & SASRec & SASRec* & +CL & 
+L2 & 
+Cos & 
+DR& 
+RD &
+CT4Rec & Improv. & Improv.*  \\
   \hline
 \multirow{4}{*}{Reorder} & HR@10   & 0.2759 & 0.2748 & 0.2768 & 0.2867 & 0.2856 & 0.2905 & 0.2857 & \textbf{0.3076} & 11.49\%& 11.94\% \\
& HR@20   & 0.3546 & 0.3392 & 0.3407 & 0.3552 & 0.3542 & 0.3558 & 0.3546 & \textbf{0.3776} &  6.49\%& 11.32\% \\
& NDCG@10 & 0.1523 & 0.1661 & 0.1856 & 0.1892 & 0.1900 & 0.1954 & 0.1899 & \textbf{0.2047} & 34.41\%& 23.24\% \\
& NDCG@20 & 0.1932 & 0.2003 & 0.2017 & 0.2064 & 0.2073 & 0.2119 & 0.2072 & \textbf{0.2224} & 15.11\%& 11.03\% \\
\hline
\multirow{4}{*}{Mask}& HR@10   & 0.2759 & 0.2748 & 0.2803 & 0.2831 & 0.2898 & 0.2892 & 0.2891 & \textbf{0.3140} & 13.81\%& 14.26\% \\
& HR@20   & 0.3546 & 0.3392 & 0.3452 & 0.3484 & 0.3553 & 0.3542 & 0.3609 & \textbf{0.3868} & 9.08\%& 14.03\% \\
& NDCG@10 & 0.1523 & 0.1661 & 0.1861 & 0.1879 & 0.1941 & 0.1948 & 0.1886 & \textbf{0.2061} & 35.33\%& 24.08\% \\
& NDCG@20 & 0.1932 & 0.2003 & 0.2025 & 0.2044 & 0.2106 & 0.2111 & 0.2067 & \textbf{0.2244} &  16.15\%& 12.03\% \\
\bottomrule
\end{tabular}}
\end{table*}

\textbf{The Effect of $\alpha$.} Figure~\ref{img:ablation} also gives the results of different $\alpha$ values. 
Considering that there are many feasible combinations of the ($\alpha$, $\beta$) grid, we temporarily remove the DR objective and only examine the influence of $\alpha$ for RD.
It can be observed that a small value of $\alpha$ can bring meaningful performance improvement.
With the increase of $\alpha$, the performance improvement further increases in which the best result is achieved when $\alpha=2.0$.
These results further confirm that the consistency regularization directly performed on the output space is very effective even when it only takes a small proportion for the final training objective.
With a further increase of $\alpha$, the model will pay more attention to the consistency of model outputs, which will dilute the original item prediction objective, resulting in worse performance (e.g., $\alpha=2.0$ v.s.,$\alpha=3.0$).

\textbf{The Influence of $\beta$.} Similar to the analysis of $\alpha$, we only study the impact of $\beta$ for each single unsupervised regularization loss (i.e., DR, cosine, and L2).
Different from $\alpha$, the DR loss has no influence on the overall performance when the value of $\beta$\textless 0.3.
This is probably because the consistency regularization on the user representations is overwhelmed and adapted when passing to the output space.
With the increase of $\beta$, DR loss gradually produces a more important role and consistently improves model performance until $\beta$\textgreater 2.0.
And also, a large $\beta$ value will force the model to focus on the consistency of user representation rather than perform the item prediction task.
For the cosine objective, different $\beta$ values have a limited influence on evaluation results.
As for the L2 loss, a large $\beta$ value will cause inferior performance.

\textbf{Analysis on Dropout Rates.} Besides the above analysis, we also study how the dropout rate affects the effectiveness of our CT4Rec since different dropout rates will lead to varying degrees of inconsistency.
As demonstrated in Figure~\ref{img:dropout}, dropout rates have a significant influence on the performance of the backbone model, and our proposed consistency training is applicable and effective when the dropout rate\textless 0.7. 
However, our proposed CT4Rec will lead to a negative effect when the dropout rate\textgreater 0.8.
We speculate that the reason behind this might be the data sparsity issue and irreconcilable inconsistency.
\subsection{Extension to Data Augmentation}
The above studies prove the effectiveness of each component of our CT4Rec and its capability of leveraging more consistency regularization signals.
We then study its generalization ability to other inconsistency scenarios.
More concretely, we utilize our consistency training method to regularize the inconsistency brought by data other than the above-mentioned model-side inconsistency, i.e., we replace dropout with data augmentation methods to create two different user representations, and the corresponded task outputs for each user so as to conduct consistency regularization.
To align with recent sequential recommendation methods based on constrastive learning, we leverage two easily implemented augmentation strategies from CL4SRec \cite{sasrec}, i.e., \textit{Reorder} and \textit{Mask}.
In doing so, we can directly compare the effects of different training objectives (besides the item prediction one for the sequential recommendation) on the augmented data, including unsupervised methods (i.e., CL, cosine, L2, DR), the supervised regularization in the output space (RD), and the combination of DR and RD (CT4Rec).
Table~\ref{tab:data_augmentation} presents the experimental results for the data augmentation setting.
We can observe that: 1) Our introduced DR objective is the most effective compared with other single methods, i.e., the consistency regularization on the representation space is the most preferable for than data augmentation scenario, which is in contrast with the observation for the dropout setting.
We speculate that the label-invariant data augmentation methods can lead to permuted and perturbed representation variants, which need more consistency regularization, while the label-invariant strategy does not deteriorate the inconsistency in the output space.
2) The combination of the consistency regularization in the representation space and the output space (CT4Rec) still performs the best with consistent and significant performance over other training objectives.

\subsection{Training and Cost Analysis}
\begin{figure}[]
    \centering
    \includegraphics[width=0.5\textwidth]{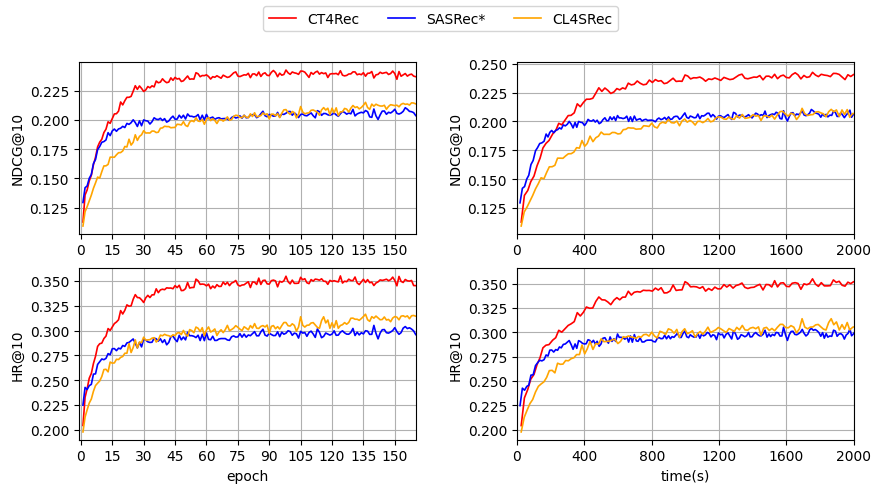}
    \caption{NDCG@10 and HR@10 curves on the valid set along with training epoch and time on the Beauty dataset.}  \label{img:time}
    \vspace{-0.25cm}
    \end{figure} 
Since our CT4Rec does not modify the model structure of the backbone model or introduce extra augmented data, we mainly analyze the changes in the training process.
We plot the curves of HR@10and NDCG@10 scores on the valid set along the training epoch number and time (seconds) for \textit{SASRec*}, \textit{CL4SRec} and our CT4Rec models, shown in Figure~\ref{img:time}.
At the early training stage, the backbone \textit{SASRec*} model converges quickly with the same training epochs (around 15 epochs) and model performance as CT4Rec, but our CT4Rec can continuously improve the performance on the well-trained \textit{SASRec*} model.
It concludes that our consistency training objectives do not lead to more training epochs on the backbone \textit{SASRec*}.
But for \textit{CL4SRec} that combines contrastive learning objective with \textit{SASRec*}, it converges with much more training epochs and only moderate performance improvement.
As for convergence time (seconds), our model indeed is slower than the backbone model since our consistency training objectives need an extra forward process (dropout) in each training step but achieves a much superior performance.
Compared with \textit{CL4SRec}, our CT4Rec is much more efficient and effective with a much better final optimum and less convergence time even when we haven't calculated the time cost of data augmentation and negative sampling for \textit{CL4SRec}.

Through the analysis, we can conclude that: 1) CT4Rec is more efficient and effective than the method based on contrastive learning even without counting its time cost of data augmentation and negative sampling; 2) CT4Rec indeed introduces extra training time for the backbone model, which can be mitigated by early stop insomuch as our CT4Rec can quickly surpass the backbone model in the early stage and with a much better final convergence performance.


\section{Extension to CTR Prediction}
\label{sec:dis}
Besides the application of CT4Rec on the matching stage, we further explore the effectiveness of our \textbf{C}onsistency \textbf{T}raining for the \textbf{CTR} prediction task denoted as CT4CTR, serving on the ranking stage. 
Concretely, for each instance $(\boldsymbol{x},y)\in\mathcal{D}$, $\boldsymbol{x}$ denotes a multi-filed feature vector input, and label \(y\in\left\{0,1\right\}\) indicates whether the user clicks the item. 
The CTR prediction is to obtain the probability \(\hat{y}\) that a user will click a certain item in a given context.

We utilize a widely used structure DeepFM in this section, which simply applies two components (i.e., FM component and deep component) and still achieves promising performance in the industry. For each input \(\boldsymbol{x}\), we forward it twice in the deep component of DeepFM with different dropouts and learn the consistency between these two sub-models. Similar to Sec.~\ref{sec:method}, two regularization methods (i.e. RD loss, DR loss) are utilized to constrain the two sub-models generated from dropout. Concretely, input \(\boldsymbol{x}\) passes the deep component twice and obtains $\hat{y}^{d_1}_{DNN}$ and $\hat{y}^{d_2}_{DNN}$ with different dropouts. Thus, with the unchanged part $\hat{y}_{FM}$, the final prediction can be defined as $\hat{y}^{d_1}$ and $\hat{y}^{d_2}$ and a two-dimensional distribution $\mathcal{P}(\boldsymbol{x}^{d_i})=(\hat{y}^{d_i},1-\hat{y}^{d_i})$ can be formulated to calculate RD loss. Meanwhile, we compare the deep representations of instances in each mini-batch,e.g., \((\boldsymbol{x_1^{d_1}}, \boldsymbol{x_2^{d_1}},\dots, \boldsymbol{x_n^{d_1}})\) and \((\boldsymbol{x_1^{d_2}}, \boldsymbol{x_2^{d_2}}, \dots, \boldsymbol{x_n^{d_2}})\). Similar to the matching task, for instance, $\boldsymbol{x_1}$, we can calculate the similarity  $sim(\boldsymbol{x_1^{d_j}},\boldsymbol{x_i^{d_j}}), i=2,3,\dots,n$, so as to obtain similarity distribution between deep representations denoted as \(\mathcal{P}_{s}(\boldsymbol{x_{1}^{d_j}})\). Then, similar to Eq. \ref{func:DR_loss}, a bidirectional KL-divergence loss to regularize the two distributions \(\mathcal{P}_{s}(\boldsymbol{x_{1}^{d_1}})\) and \(\mathcal{P}_{s}(\boldsymbol{x_{1}^{d_2}})\) can be denoted as DR loss.

We compare CT4CTR with: (1) LR \cite{LR}, a simple baseline model for CTR prediction, which only models the linear combination of raw features. (2) FM \cite{FM}. Since LR fails to capture non-linear feature interactions, the factorization machine (FM) has been proposed to model second-order feature interactions. It is notable that FM only has a linear time complexity in terms of the number of features. (3) Wide\&Deep\cite{WideDeep}. With the development of deep models, Google achieves great improvement by combining a wide (or shallow) network and a deep network. This is a general learning framework that can achieve the advantages of both wide networks and deep networks. (4) DeepFM~\cite{deepfm}, which extends Wide\&Deep by substituting LR with FM to precisely model second-order feature interactions. (5) xDeepFM\cite{xdeepfm}, which captures high-order interactions by its core module, Compressed Interaction Network (CIN). CIN takes an outer product of a stacked feature matrix in a vector-wise way. (6) AutoInt\cite{Autoint}, which automatically models the high-order interactions of input features by using self-attention networks.

\begin{table}[t]
    \caption{Dataset statistics of two corpora from real-world recommendation system in WeChat platform.}
    \begin{tabular}{ccccc}
    \toprule
         Dataset & \#Instances & \#Fields & \#Features  \\
         \hline
         Wechat-Video & 41M & 39 & 40.28M   \\
         Wechat-Article & 138M & 25 & 21.52M   \\
\bottomrule     
    \end{tabular}
    \label{ctr}
     \vspace{-0.3cm}
\end{table}
To evaluate the performance of CT4CTR, we build two private industrial datasets from the WeChat ecosystem: Wechat-Video and Wechat-Article, and both of them are collected from a real-world recommendation scenario, i.e., WeChat Subscriptions. We choose them because they are sampled from real click logs in production, and both have tens of millions of samples, making the results meaningful to industrial practitioners. We split Train and Test sets in chronological order, and Table \ref{ctr} summarizes the detailed statistics.
\begin{table}[th]
\caption{Offline performance of CT4CTR in the CTR prediction task, with DeepFM as backbone.}
\begin{tabular}{ll|ll}
\hline
\multicolumn{2}{l|}{WeChat-Video} & \multicolumn{2}{l}{WeChat-Article} \\ \hline
Model             & AUC           & Model              & AUC           \\ \hline
LR                & 0.7569        & LR                 & 0.7401        \\
FM                & 0.7608        & FM                 & 0.7465        \\
Wide\&Deep        & 0.7695        & Wide\&Deep         & 0.7538        \\
DeepFM            & 0.7719        & DeepFM             & 0.7552        \\
AutoInt           & 0.7703        & AutoInt            & 0.7519        \\
xDeepFM           & 0.7738        & xDeepFM            & 0.7560        \\
CT4CTR            & 0.7766        & CT4CTR             & 0.7593        \\ \hline
\end{tabular}
\label{auc}
\end{table}

As shown in Table \ref{auc}, our proposed CT4CTR outperforms all other baseline methods in CTR prediction on AUC, including multiple state-of-the-art CTR prediction solutions, on two large industrial corpora. 
Besides, we have deployed CT4CTR on a real-world recommendation system that serves nearly one billion users with dramatically high online machine costs.
In the online A/B test, we achieve +1.203\% on the Video scenario and +2.341\% on the Article scenario, as for the CTR metric. 
Moreover, it is a remarkable success that CT4CTR can enhance online performance without involving extra computation costs and machine costs, which is essential for online serving.
Besides, the improvement of such a mature system that already has stable and advanced online models is extremely challenging, which further proves the effectiveness of CT4CTR.

\begin{table}[th]
    \caption{Performance improvement of online A/B test.}
    \centering
    \begin{tabular}{c|cc}
        \toprule
         Scenario   & Video  & Article  \\
         \hline
         CTR   & +1.203\%  & +2.341\%  \\
        \bottomrule     
    \end{tabular} 
    \label{auc2}
    \vspace{-0.3cm}
\end{table}

 





\section{Conclusion and Future Work}\label{sec:conclusion}
In this paper, we proposed a simple yet very effective consistency training method for the sequential recommendation task, namely CT4Rec, which only involves two bidirectional KL losses.
We first introduce a top-performed regularization in the output space by minimizing the bidirectional KL loss of two different outputs.
We then design a novel consistency training term in the representation space by minimizing the distributed probability of two user representations.
Extensive experiments and analysis demonstrate its effectiveness, efficiency, generalization ability, and compatibility.
Besides experiments on the recall task in recommendation systems, further exploration reveals that the introduced consistency training strategies (i.e., DR and RD) are still very effective for the CTR prediction task.
In the near future, we will thoroughly study the pros and cons of consistency training for the CTR task.

\bibliographystyle{ACM-Reference-Format}
\balance
\bibliography{tkde2020}



\clearpage
\appendix
\section{Appendix}
\subsection{Model structure of CT4CTR}
\begin{figure}[htbp]
    \includegraphics[width=0.5\textwidth]{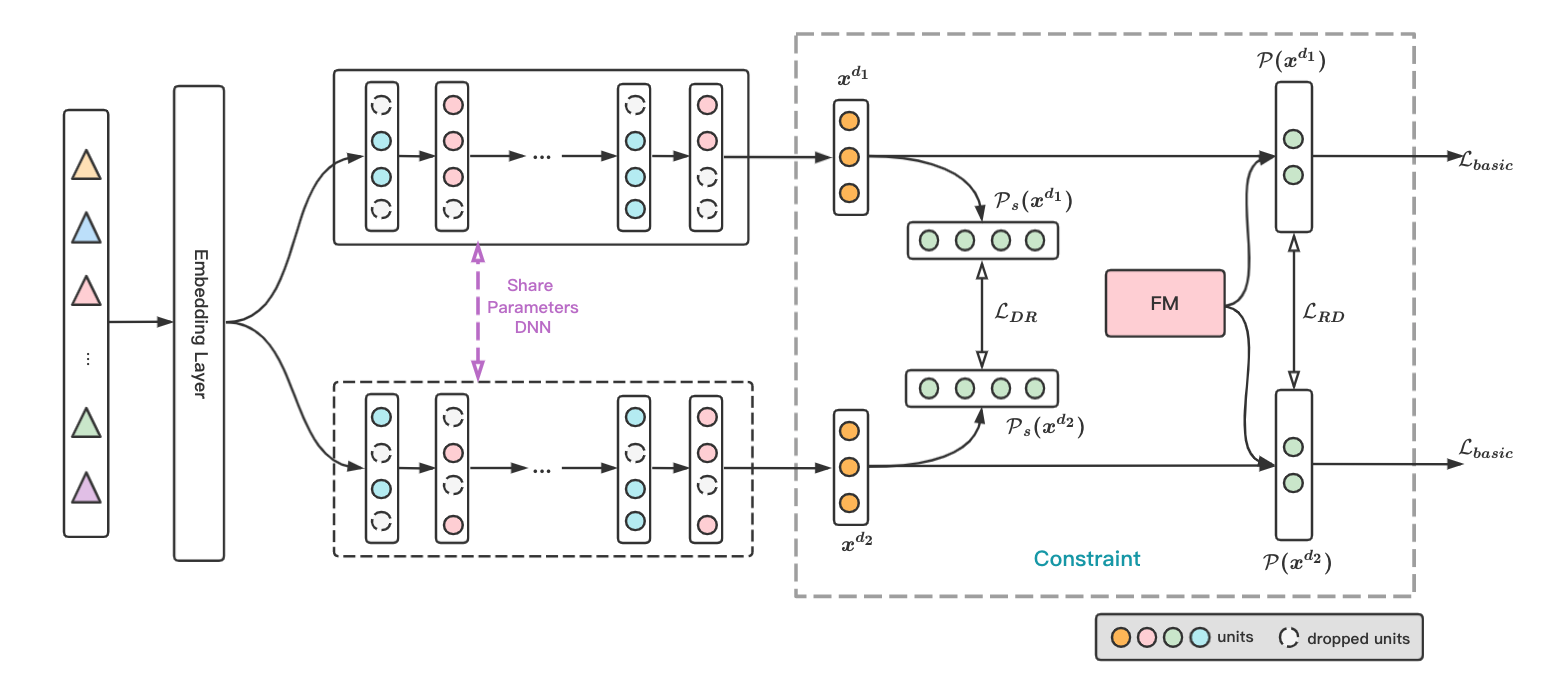}
    \caption{Model structure of CT4CTR.}  \label{CT4Rec-rank}
    \end{figure}

The model structure of CT4CTR is shown in Figure~\ref{CT4Rec-rank}. CT4CTR takes user features and item features as input and outputs click probabilities for each user and item pair in the ranking stage of recommendation. For the deep component of DeepFM, input features are transformed into vector representations via the embedding layer and deep neural network with different hidden dropout masks. In addition, Distributed Regularization Loss and Regularized Dropout Loss are introduced to restrain these representations generated by different dropout masks.

\subsection{Training Algorithm}\label{sec:app1}
\begin{algorithm}
\caption{CT4Rec algorithm}\label{alg:CT4Rec}
\renewcommand{\algorithmicrequire}{\textbf{Input:}}
\renewcommand{\algorithmicensure}{\textbf{Output:}}
\begin{algorithmic}[1]
    \Require Training data  $\mathcal{D}=\left\{s_{u_i,t}\right\}^N_{i=1} $ 
    \Ensure model parameters $\omega$ 
    \State Initialization model with parameters $\omega$
    \While{not converged}
		\State  $s_{u_i,t} \sim \mathcal{D} $
		\State  $\boldsymbol{s_{u_i,t}^{d_1}} \leftarrow f(s_{u_i,t};\omega)$ with first dropout 
		\State  $\boldsymbol{s_{u_i,t}^{d_2}} \leftarrow f(s_{u_i,t};\omega)$ with second dropout 
		\State  $  g  \leftarrow \bigtriangledown_\omega \mathcal{L}_{final}(\boldsymbol{s_{u_i,t}^{d_1}},\boldsymbol{s_{u_i,t}^{d_2}};\omega)$
		\State  $\omega \leftarrow GradientUpdate(\omega,g) $ 
    \EndWhile
\end{algorithmic}
\end{algorithm}
The whole training process of CT4Rec is presented in Algorithm \ref{alg:CT4Rec}. As shown in Line 3-5, we obtain two user representations \(\boldsymbol{s_{u_i,t}^{d_1}}\) and \(\boldsymbol{s_{u_i,t}^{d_2}}\) by going forward the model twice for each user sequence \(s_{u_i,t}\). Line 6-7 calculate the \(\mathcal{L}_{final}\) according to the loss function \ref{func:final_loss}, and update the model parameters. The training process will continue until convergence.

\subsection{More Backbone}
To verify the universal influence of CT4Rec, we also apply it to GRU4Rec~\cite{GRU4Rec} denoted as CT4Rec$^G$ and evaluate its performance on four datasets. Table~\ref{tab:gru4rec} shows that CT4Rec brings significant improvement compared with the original GRU4Rec on all datasets and all HR\({@k}\) and NDCG\({@k}\) scores (relative improvements ranging from 3.48\% to 11.51\%), which indicates that our method can be widely applied to different model structures and achieve enhancement.
\begin{table}[th]
    \caption{Performance of CT4Rec with GRU4Rec as  backbone}
    \centering
    \small
    \begin{tabular}{lllll}
    \hline 
    Datasets   & Metric & GRU4Rec & CT4Rec$^G$ & Improv.      \\
    \hline 
    Beauty & HR@5    & 0.1149~ & 0.1247~          & 8.53\%   \\
           & HR@10   & 0.1574~ & 0.1665~          & 5.78\%   \\
           & HR@20   & 0.2157~ & 0.2232~          & 3.48\%   \\
           & NDCG@5  & 0.0851~ & 0.0923~          & 8.46\%   \\
           & NDCG@10 & 0.0987~ & 0.1057~          & 7.09\%   \\
           & NDCG@20 & 0.1133~ & 0.1200~          & 5.91\%   \\
           \hline 
    Sport  & HR@5    & 0.0997~ & 0.1057~          & 6.02\%   \\
           & HR@10   & 0.1558~ & 0.1665~          & 6.87\%   \\
           & HR@20   & 0.2356~ & 0.2504~          & 6.28\%   \\
           & NDCG@5  & 0.0656~ & 0.0699~          & 6.55\%   \\
           & NDCG@10 & 0.0836~ & 0.0894~          & 6.94\%   \\
           & NDCG@20 & 0.1037~ & 0.1105~          & 6.56\%   \\
           \hline 
    Yelp   & HR@5    & 0.1460~ & 0.1628~          & 11.51\%  \\
           & HR@10   & 0.2570~ & 0.2817~          & 9.61\%   \\
           & HR@20   & 0.4335~ & 0.4601~          & 6.14\%   \\
           & NDCG@5  & 0.0903~ & 0.1011~          & 11.96\%  \\
           & NDCG@10 & 0.1259~ & 0.1392~          & 10.56\%  \\
           & NDCG@20 & 0.1702~ & 0.1840~          & 8.11\%  \\
            \hline 
    WeChat & HR@5    & 0.1833  & 0.1946           & 6.61\%   \\
       & HR@10   & 0.2235  & 0.2378           & 6.40\%   \\
       & HR@20   & 0.2891  & 0.3042           & 5.22\%   \\
       & NDCG@5  & 0.1267  & 0.1369           & 8.05\%   \\
       & NDCG@10 & 0.1450  & 0.1581           & 9.03\%   \\
       & NDCG@20 & 0.1694  & 0.1804           & 6.49\%  \\
           \hline 
    \end{tabular}
    \label{tab:gru4rec}
\end{table}

\subsection{More Experimental Results}\label{sec:app2}
Here, we present the experimental results of extra two runs with different random seeds.
\begin{table*}[th]
\caption{The second run of baselines and our proposed CT4Rec on four offline datasets, where `*' refers to modifying the original binary cross-entropy loss in SASRec with the training objective in recent baselines, e.g., CLRec, CL4SRec, StackRec. Our CT4Rec is implemented on SASRec*. 
\textit{Improv.} and \textit{Improv.*} refer to the relative improvement of CT4Rec over SASRec and SASRec*, respectively. Except for the random seed, the other settings are the same as Table~\ref{tb2}.}
\renewcommand\arraystretch{0.88}
\resizebox{\textwidth}{!}{
\begin{tabular}{lllllllllllll}
\toprule  
Datasets   & Metric & SASRec & SASRec* & GRU4Rec & BERT4Rec & TiSASRec & StackRec & CLRec  & CL4SRec & CT4Rec  & Improv. & Improv.* \\
\hline
                         & HR@5                                & 0.2127 & 0.2162  & 0.1184  & 0.0835   & 0.2087   & 0.1685   & 0.1528 & 0.2284  & \textbf{0.2565} & 20.59\% & 18.64\%  \\
                         & HR@10                               & 0.2805 & 0.2707  & 0.1592  & 0.1323   & 0.2805   & 0.2167   & 0.1872 & 0.2894  & \textbf{0.3211} & 14.47\% & 18.62\%  \\
                         & HR@20                               & 0.3583 & 0.3340  & 0.2157  & 0.1997   & 0.3569   & 0.2779   & 0.2288 & 0.3606  & \textbf{0.3915} & 9.27\%  & 17.22\%  \\
                         & NDCG@5 & 0.1521 & 0.1639  & 0.0873  & 0.0550   & 0.1450   & 0.1272   & 0.1198 & 0.1704  & \textbf{0.1927} & 26.69\% & 17.57\%  \\
                         & NDCG@10                             & 0.1740 & 0.1815  & 0.1004  & 0.0707   & 0.1682   & 0.1427   & 0.1309 & 0.1901  & \textbf{0.2136} & 22.76\% & 17.69\%  \\
\multirow{-7}{*}{Beauty} & NDCG@20                             & 0.1937 & 0.1975  & 0.1147  & 0.0876   & 0.1875   & 0.1581   & 0.1415 & 0.2081  & \textbf{0.2314} & 19.46\% & 17.16\%  \\
\hline
                         & HR@5                                & 0.1918 & 0.1963  & 0.0978  & 0.0738   & 0.1716   & 0.1321   & 0.1529 & 0.2052  & \textbf{0.2226} & 16.06\% & 13.40\%  \\
                         & HR@10                               & 0.2753 & 0.2684  & 0.1543  & 0.1250   & 0.2457   & 0.1964   & 0.2187 & 0.2798  & \textbf{0.3026} & 9.92\%  & 12.74\%  \\
                         & HR@20                               & 0.3727 & 0.3550  & 0.2336  & 0.2048   & 0.3342   & 0.2797   & 0.3080 & 0.3687  & \textbf{0.3964} & 6.36\%  & 11.66\%  \\
                         & NDCG@5 & 0.1302 & 0.1393  & 0.0651  & 0.0481   & 0.1166   & 0.0907   & 0.1064 & 0.1457  & \textbf{0.1573} & 20.81\% & 12.92\%  \\
                         & NDCG@10                             & 0.1571 & 0.1626  & 0.0832  & 0.0645   & 0.1405   & 0.1111   & 0.1276 & 0.1697  & \textbf{0.1831} & 16.55\% & 12.61\%  \\
\multirow{-7}{*}{Sport}  & NDCG@20                             & 0.1816 & 0.1844  & 0.1032  & 0.0846   & 0.1627   & 0.1321   & 0.1501 & 0.1921  & \textbf{0.2067} & 13.82\% & 12.09\%  \\
\hline
                         & HR@5                                & 0.2826 & 0.3180  & 0.1463  & 0.1518   & 0.2938   & 0.2204   & 0.2511 & 0.3159  & \textbf{0.3485} & 23.32\% & 9.59\%   \\
                         & HR@10                               & 0.4252 & 0.4421  & 0.2589  & 0.2557   & 0.4225   & 0.3298   & 0.3855 & 0.4416  & \textbf{0.4807} & 13.05\% & 8.73\%   \\
                         & HR@20                               & 0.5991 & 0.5941  & 0.4345  & 0.4202   & 0.5815   & 0.4620   & 0.5648 & 0.5967  & \textbf{0.6342} & 5.86\%  & 6.75\%   \\
                         & NDCG@5 & 0.1870 & 0.2257  & 0.0906  & 0.0954   & 0.2020   & 0.1479   & 0.1680 & 0.2212  & \textbf{0.2459} & 31.50\% & 8.95\%   \\
                         & NDCG@10                             & 0.2329 & 0.2657  & 0.1268  & 0.1287   & 0.2434   & 0.1831   & 0.2113 & 0.2618  & \textbf{0.2886} & 23.92\% & 8.62\%   \\
\multirow{-7}{*}{Yelp}   & NDCG@20                             & 0.2767 & 0.3039  & 0.1708  & 0.1700   & 0.2835   & 0.2165   & 0.2565 & 0.3009  & \textbf{0.3273} & 18.29\% & 7.70\%  \\
\hline
                         & HR@5                                & 0.2763 & 0.3067  & 0.1825  & 0.1924   & 0.3162   & 0.2981   & 0.2868 & 0.3103  & \textbf{0.3416} & 23.63\% & 11.38\%  \\
                         & HR@10                               & 0.4113 & 0.4356  & 0.2238  & 0.2261   & 0.4401   & 0.4165   & 0.3993 & 0.4504  & \textbf{0.4854} & 18.02\% & 11.43\%  \\
                         & HR@20                               & 0.5274 & 0.5481  & 0.2897  & 0.2926   & 0.5538   & 0.5180   & 0.4871 & 0.5511  & \textbf{0.5981} & 13.41\% & 9.12\%   \\
                         & NDCG@5 & 0.1961 & 0.2152  & 0.1287  & 0.1252   & 0.2060   & 0.2066   & 0.1947 & 0.2207  & \textbf{0.2373} & 21.01\% & 10.27\%  \\
                         & NDCG@10                             & 0.2347 & 0.2762  & 0.1472  & 0.1416   & 0.2583   & 0.2583   & 0.2379 & 0.2817  & \textbf{0.3070} & 30.81\% & 11.15\%  \\
\multirow{-7}{*}{WeChat} & NDCG@20                             & 0.2882 & 0.3030  & 0.1683  & 0.1547   & 0.2977   & 0.2806   & 0.2659 & 0.3083  & \textbf{0.3335} & 15.72\% & 10.07\% \\
\bottomrule
\end{tabular}}
\end{table*}
\begin{table*}[th]
\caption{The third run of baselines and our proposed CT4Rec on four offline datasets, where `*' refers to modifying the original binary cross-entropy loss in SASRec with the training objective in recent baselines, e.g., CLRec, CL4SRec, StackRec. Our CT4Rec is implemented on SASRec*. 
\textit{Improv.} and \textit{Improv.*} refer to the relative improvement of CT4Rec over SASRec and SASRec*, respectively. Except for the random seed, the other settings are the same as Table~\ref{tb2}.}
\renewcommand\arraystretch{0.88}
\resizebox{\textwidth}{!}{
\begin{tabular}{lllllllllllll}
\toprule  
Datasets   & Metric & SASRec & SASRec* & GRU4Rec & BERT4Rec & TiSASRec & StackRec & CLRec  & CL4SRec & CT4Rec  & Improv. & Improv.* \\
\hline
                         & HR@5                                & 0.2087                     & 0.2198                      & 0.1149                      & 0.0840   & 0.2062                       & 0.1697   & 0.1528                    & 0.2240                      & \textbf{0.2576}            & 23.43\% & 17.20\%  \\
                         & HR@10                               & 0.2758                     & 0.2740                      & 0.1574                      & 0.1319   & 0.2768                       & 0.2157   & 0.1853                    & 0.2842                      & \textbf{0.3208}            & 16.32\% & 17.08\%  \\
                         & HR@20                               & 0.3517                     & 0.3360                      & 0.2157                      & 0.2025   & 0.3522                       & 0.2764   & 0.2255                    & 0.3558                      & \textbf{0.3911}            & 11.20\% & 16.40\%  \\
                         & NDCG@5 & 0.1482                     & 0.1674                      & 0.0851                      & 0.0550   & 0.1441                       & 0.1272   & 0.1198                    & 0.1679                      & \textbf{0.1933}            & 30.43\% & 15.47\%  \\
                         & NDCG@10                             & 0.1699                     & 0.1849                      & 0.0987                      & 0.0704   & 0.1668                       & 0.1420   & 0.1303                    & 0.1873                      & \textbf{0.2138}            & 25.84\% & 15.63\%  \\
\multirow{-7}{*}{Beauty} & NDCG@20                             & 0.1890                     & 0.2005                      & 0.1133                      & 0.0880   & 0.1859                       & 0.1573   & 0.1404                    & 0.2054                      & \textbf{0.2315}            & 22.49\% & 15.46\%  \\
\hline
                         & HR@5                                & 0.1908                     & 0.1968                      & 0.0997                      & 0.0781   & 0.1721                       & 0.1334   & 0.1543                    & 0.2082                      & \textbf{0.2198}            & 15.20\% & 11.69\%  \\
                         & HR@10                               & 0.2734                     & 0.2680                      & 0.1558                      & 0.1294   & 0.2447                       & 0.1979   & 0.2215                    & 0.2832                      & \textbf{0.2975}            & 8.81\%  & 11.01\%  \\
                         & HR@20                               & 0.3733                     & 0.3546                      & 0.2356                      & 0.2090   & 0.3328                       & 0.2802   & 0.3106                    & 0.3731                      & \textbf{0.3915}            & 4.88\%  & 10.41\%  \\
                         & NDCG@5 & 0.1308                     & 0.1409                      & 0.0656                      & 0.0502   & 0.1169                       & 0.0908   & 0.1077                    & 0.1481                      & \textbf{0.1556}            & 18.96\% & 10.43\%  \\
                         & NDCG@10                             & 0.1574                     & 0.1638                      & 0.0836                      & 0.0666   & 0.1403                       & 0.1115   & 0.1293                    & 0.1723                      & \textbf{0.1807}            & 14.80\% & 10.32\%  \\
\multirow{-7}{*}{Sport}  & NDCG@20                             & 0.1825                     & 0.1856                      & 0.1037                      & 0.0866   & 0.1625                       & 0.1322   & 0.1517                    & 0.1949                      & \textbf{0.2044}            & 12.00\% & 10.13\%  \\
\hline
                         & HR@5                                & 0.2887                     & 0.3194                      & 0.1460                      & 0.1522   & 0.2976                       & 0.2202   & 0.2573                    & 0.3214                      & \textbf{0.3475}            & 20.37\% & 8.80\%   \\
                         & HR@10                               & 0.4313                     & 0.4471                      & 0.2570                      & 0.2559   & 0.4294                       & 0.3295   & 0.3911                    & 0.4489                      & \textbf{0.4803}            & 11.36\% & 7.43\%   \\
                         & HR@20                               & 0.6023                     & 0.6011                      & 0.4335                      & 0.4227   & 0.5857                       & 0.4628   & 0.5649                    & 0.6013                      & \textbf{0.6326}            & 5.03\%  & 5.24\%   \\
                         & NDCG@5 & 0.1922                     & 0.2264                      & 0.0903                      & 0.0962   & 0.2045                       & 0.1476   & 0.1725                    & 0.2250                      & \textbf{0.2458}            & 27.89\% & 8.57\%   \\
                         & NDCG@10                             & 0.2381                     & 0.2675                      & 0.1259                      & 0.1294   & 0.2470                       & 0.1828   & 0.2155                    & 0.2661                      & \textbf{0.2886}            & 21.21\% & 7.89\%   \\
\multirow{-7}{*}{Yelp}   & NDCG@20                             & 0.2812                     & 0.3064                      & 0.1702                      & 0.1712   & 0.2865                       & 0.2165   & 0.2592                    & 0.3045                      & \textbf{0.3270}            & 16.29\% & 6.72\%  \\
\hline
                         & HR@5                                & 0.2726 & 0.3073  & 0.1842  & 0.1937   & 0.3214   & 0.2958   & 0.2864 & 0.3112  & \textbf{0.3409} & 25.06\% & 10.93\%  \\
                         & HR@10                               & 0.4091 & 0.4361  & 0.2261  & 0.2253   & 0.4417   & 0.4181   & 0.4003 & 0.4517  & \textbf{0.4852} & 18.60\% & 11.26\%  \\
                         & HR@20                               & 0.5280 & 0.5472  & 0.2891  & 0.2926   & 0.5554   & 0.5208   & 0.4868 & 0.5502  & \textbf{0.5965} & 12.97\% & 9.01\%   \\
                         & NDCG@5 & 0.1932 & 0.2143  & 0.1277  & 0.1271   & 0.2051   & 0.2061   & 0.1953 & 0.2175  & \textbf{0.2353} & 21.79\% & 9.80\%   \\
                         & NDCG@10                             & 0.2352 & 0.2747  & 0.1457  & 0.1422   & 0.2631   & 0.2562   & 0.2362 & 0.2819  & \textbf{0.3067} & 30.40\% & 11.65\%  \\
\multirow{-7}{*}{WeChat} & NDCG@20                             & 0.2834 & 0.3007  & 0.1706  & 0.1542   & 0.2973   & 0.2803   & 0.2647 & 0.3077  & \textbf{0.3301} & 16.48\% & 9.78\% \\
\bottomrule
\end{tabular}}
\end{table*}

\end{document}